\documentclass[prl,twocolumn,nobalancelastpage]{revtex4}
\usepackage{graphicx}
\usepackage{times}
\begin{document}
{\bf Kortus et al.  Reply:} Samuely et al. \cite{SamComment} make
the strong claim that our letter \cite{kortus} is 'contradicting
several established experimental results'. As we will show below
this is not justified and the claims from
\cite{SamComment} are based on a misinterpretation of our results.

To clarify the case, we reproduce Fig.~1 from
\cite{SamComment} together with our prediction of the
$T_c$ dependence of the gaps without interband scattering (Fig.2
from \cite{kortus}). Our results without interband scattering
reproduce nicely the new data points, which partially were not
available at submission of \cite{kortus}. The
straightforward conclusion is that these samples
have very small interband scattering.

In our letter we never claimed that the merging
of the $\sigma$- and $\pi$-gaps observed in \cite{gonnelli} is an
universal behavior for C doped MgB$_2$ samples, and
we emphasized this view by expressions like 'indicate for
the first time' \cite{kortus} and referencing several
other works not showing this behavior. Obviously, the
behavior will depend on the defects which may cause interband
scattering in MgB$_2$, which can be very different for
samples produced by different techniques in different places. This
becomes immediately clear from our Fig.~1 from our letter \cite{kortus}, 
if all samples with the same doping concentration would have the same
interband scattering rate their critical temperature $T_c$ should
be the same. This is clearly not the case, in particular for
higher doping concentrations.

Our model allows to describe, as far as we aware of, all
available experimental data by simply varying the interband
scattering rate. The model does not depend on or require 
any assumption on any particular experiment.
Further, our letter \cite{kortus}
does not contain any explicit statement,
as implied by \cite{SamComment}, that
the two gaps have to merge in C-doped samples at 10\% doping. The
doping concentration at which the two gaps will merge, depends on
the magnitude of interband scattering. Our abstract and
conclusion of \cite{kortus} clearly state that the compensating
effect of band filling and interband scattering shifts the merging
of the the gaps to higher doping concentrations which is entirely
correct.
The reason why we emphasized the data from
\cite{gonnelli} is, that if one observes the merging of the two
gaps (or even just a constant $\pi$-gap as function of doping \cite{pono})
interband scattering has to be taken into account in
addition to the band filling effect. The experimental results of
\cite{gonnelli} are reproduced by a very simple assumption for the
magnitude of the interband scattering, which may be
correct only for this particular set of samples.

Erwin and Mazin \cite{erwin} show, that a single C impurity
replacing B does not relax out of the boron plane due to symmetry
and induces relatively large in plane relaxation of the
neighboring boron atoms. Using some approximations the authors
relate the out of plane displacement to the interband scattering,
which would be zero for such a C impurity. This is only 
at a first superficial sight in contrast to our results. 
We already pointed out that $T_c$ varies for samples with the 
same doping concentration, which clearly indicates that the C-doping  
is not directly generating the interband scattering. However,
C impurities in the boron plane may induce other defects which
cause interband scattering. Other possibilities which may be
relevant at higher doping concentrations include e.g.\ the
formation of pairs or clusters of carbon which may allow for
buckling out of the boron plane. Therefore, if the concentration
of defects which cause interband scattering is proportional to the
C doping, the simple linear assumption used to reproduce the
results of \cite{gonnelli} would still hold.

\begin{figure} [t]
\begin{center}
\includegraphics[width=\linewidth,clip=true]{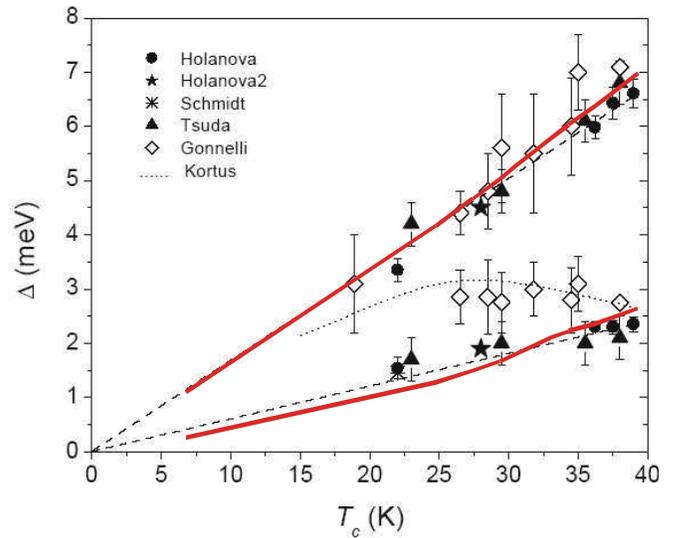}
\end{center}
\caption{The Figure reproduces Fig.~1 from \protect\cite{SamComment} together
with our line (solid) without interband scattering from Fig.~2 from our letter \protect\cite{kortus}.}
\end{figure}

In summary, we are confident that our model \cite{kortus} captures
the essential features to understand the doping behavior in
MgB$_2$, including the new data presented in \cite{SamComment}. It
rather appears that Samuely et al. \cite{SamComment} question the
experimental results by Gonnelli et al. \cite{gonnelli}, than our
model which does not depend on any assumption of the behavior of
any particular experiment. Even if the data reported in
\cite{gonnelli} would be questionable, for which we see at the
moment no good reason, our theoretical description of the
doping behavior of the two gaps and $T_c$ would still be correct.
In that case experimentalists should see our work as a challenge
to produce samples with controlled interband  scattering rates,
e.g.\ following the suggestion by Erwin and Mazin \cite{erwin} of
codoping Al and Na, in order to obtain the predicted
merging of the two superconducting gaps.

%\vspace*{1ex}
\noindent
J. Kortus$^{1,2}$, O.V. Dolgov$^{2}$, R.K.\ Kremer$^{2}$
and A.A.\ Golubov$^3$ \\
$^{1}$ IPCMS-GMI Strasbourg, France\\
$^{2}$ MPI f{\"{u}}r Festk{\"{o}}rperforschung Stuttgart, Germany\\
$^{3}$ MESA+ Research Institute and University of Twente, The Netherlands

\end{document}